\documentstyle[aps,graphicx]{revtex}

\begin{document}
\preprint{\vbox{\hbox{UCD-HEP-99-8} }}
\draft
\title {Diphoton signals for Low Scale Gravity in Extra Dimensions}
\author{Kingman Cheung}
\address{ 
Department of Physics, University of California, Davis, 
CA 95616 USA}
\date{\today}
\maketitle
\begin{abstract}
Gravity can become strong at the TeV scale in the theory of extra dimensions.
An effective Lagrangian can be used to describe the gravitational interactions
below a cut-off scale.  In this work, we study the diphoton production
in $\gamma\gamma$, $p\bar p$, and $e^+ e^-$  collisions in the model
of low scale gravity.
Since in the standard model photon-photon scattering only occurs via box
diagrams, the cross section is highly suppressed.  Thus, photon-photon
scattering opens an interesting opportunity for studying the new gravity
interaction, which allows tree-level photon couplings.
In addition, we also examine the diphoton production at hadronic and 
$e^+ e^-$  colliders.  We derive the limits on the cut-off scale 
from the available diphoton data and also estimate the sensitivity
reach in Run II at the Tevaton and at the future linear $e^+ e^-$
colliders.
\end{abstract}

\section{Introduction}

Recent advances in string theories suggest that  a special
11-dimension theory (dubbed as M theory) \cite{mtheory} may be the theory of 
everything.  Impacts of M theory on our present world can be studied with
compactification of the 11 dimensions down to our $3+1$ dimensions.  The 
path of compactification is, however, not unique.  
In this multi-dimensional world, the standard model particles live on a 
brane ($3+1$ dim)
while there are other fields, like gravity and super Yang-Mill fields, 
live in the bulk.  
The scale at which the extra dimensions are felt is unknown, anywhere from
TeV to Planck scale.  Recent studies \cite{gut} show that if this scale is
of order TeV and there are gauge and fermion fields living in the bulk 
that correspond to the Kaluza-Klein excitations of the gauge and fermion fields
of the SM, early unification of gauge couplings can be realized below 
or even much below the original GUT scale.  This is possible because the
extra matters in the bulk accelerate the RGE running 
of the gauge couplings, which then
change from logarithmic evolution to power evolution.  
Supersymmetry model building is also an active area in the framework of
extra dimensions \cite{msusy}.
Apart from the above, radical ideas like TeV scale string theories
were also proposed \cite{tevstring}. 

Inspired by string theories a simple but probably workable solution to
the gauge hierarchy was recently 
proposed by Arkani-Hamed, Dimopoulos and Dvali (ADD)
\cite{theory}.  They assumed the space is $4+n$ dimensional, with the
SM particles living on a brane.  While the electromagnetic, strong,
and weak forces are confined to this brane, gravity can propagate 
in the extra dimensions.  To solve the gauge hierarchy problem they
proposed the ``new'' Planck scale $M_S$ is of the order of 
TeV in this picture with
the extra dimensions of a very large size $R$.
The usual Planck scale 
$M_G=1/\sqrt{G_N} \sim 1.22 \times 10^{19}$ GeV is related to this effective
Planck scale $M_S$ using the Gauss's law:
\begin{equation}
R^n \, M_S^{n+2} \sim M_G^2 \;.
\end{equation}
For $n=1$ it gives a large value for $R$, which is already ruled out 
by gravitational
experiments.  On the other hand, $n=2$ gives $R \alt 1$ mm, which is 
in the margin beyond the reach of present gravitational experiments.
  
The graviton including its excitations in the extra dimensions 
can couple to the SM particles on the brane with an effective strength of
$1/M_S$ (instead of $1/M_G$) after summing the effect 
of all excitations collectively, and thus the gravitation interaction
becomes comparable in strength to weak interaction at TeV scale.  Hence, it
can give rise to a number of phenomenological activities testable at existing
and future colliders \cite{wells,shrock,mira,han,joanne,tom1,id1,id2,bere,desh,grae,nath,ours,tom2,atwood,yuan,id3,id4}.  
So far, studies show that there
are two categories of signals: direct and indirect.  The indirect signal
refers to exchanges of gravitons in the intermediate states, while direct
refers to production or associated production of gravitons in the final 
state \cite{wells,mira,han,ours,atwood,yuan}.  
Indirect signals include fermion pair,
gauge boson pair production, correction to precision variables, etc. 
\cite{wells,shrock,han,joanne,tom1,id1,id2,bere,desh,grae,nath,tom2,id3,id4}.
There are also other astrophysical and cosmological signatures and 
constraints \cite{others}.

Processes that only occur via loop diagrams in the SM are especially
interesting if the low scale gravity allows tree-level interactions.
In the SM, the lowest order photon-photon scattering can only take place
via box diagrams of order $\alpha^2$ (on amplitude level) 
\cite{photon} and,  therefore, is highly
suppressed.  Thus, photon-photon scattering opens an interesting door
for any tree-level photon interactions. Even if such new interactions are much
weaker than the electroweak strength, these tree-level diagrams
are only of order $\alpha_{\rm new}$.  It stands a good chance that
these new interactions can beat the standard model.  
In the framework of ADD, photons can scatter via exchanges of spin-2 
gravitons in $s$-, $t$-, and $u$-channels and the most important is that
the coupling strength can be as large as the electroweak strength.
In this work, we shall study the photon-photon scattering $\gamma\gamma
\to \gamma\gamma$  and demonstrate
that it provides a unique channel to identify the low scale gravity
interactions.  
Other interesting processes of the same category is $\gamma\gamma
\to \nu \bar \nu$ and the cross-channel, $\gamma \nu \to \gamma \nu$, 
both of which
do not have any tree-level contributions in the SM \cite{duane0}.  We shall
not pursue these two further in this paper.

Similarly, a pair of gluons can scatter into gluons or photons via
exchanges of gravitons, the latter of which is our attention at hadron
colliders.  The lowest order $gg\to \gamma\gamma$ scattering occurs via a 
$s$-channel exchange of graviton in the low scale gravity model whereas 
it has to be
via box diagrams in the SM.  Thus, the new gluon scattering will give rise 
to anomalous diphoton production, in addition to 
the $q \bar q \to G \to \gamma\gamma$ channel, at hadron colliders.
However, the tree-level SM $q \bar q \to \gamma \gamma$ presents a large
irreducible background, not to mention the jet-fake background.
This makes the diphoton production at hadron colliders not as attractive 
as in $\gamma\gamma$ and $e^+ e^-$ colliders as a probe to the low scale
gravity model.  
For completeness we also study the diphoton production at $e^+ e^-$ colliders.

The organization of the paper is as follows.  In the next section, we 
compare the photon-photon scattering cross section between the SM and
the low scale gravity.  In Sec. III, we calculate diphoton production 
at the Tevatron and obtain the present limit on the cut-off scale $M_S$ using 
the diphoton data, and then estimate the sensitivity reach at the Run II. 
In Sec. IV, we repeat the same exercise at $e^+ e^-$ colliders
and obtain the limits using the diphoton data from LEPII, and
estimate the sensitivity reach at the future linear $e^+ e^-$ colliders.
We shall then conclude in Sec. V.

\section{Photon-photon Scattering}

We concentrate on the spin-2 component of the Kaluza-Klein (KK) 
states, which are the excited modes of  graviton in the extra dimensions.
The spin-0 component has a coupling to the gauge boson 
proportional to the mass of the gauge
boson in the unitary gauge, which means it has a zero
coupling to photons.  We follow the convention  in Ref.\cite{han}.
There are three contributing Feynman diagrams for the process $\gamma\gamma
\to \gamma\gamma$ in the $s$-, $t$-, and $u$-channels.
The amplitudes for $\gamma(p_1) \gamma(p_2) \to \gamma(k_1) \gamma(k_2)$  
are given by:
\begin{eqnarray}
i{\cal M}_1 &=& - \frac{\kappa^2}{8} D(t)\, B^{\mu\nu,\mu'\nu'}(p_1-k_1)\,
\epsilon^\rho(p_1) \, \epsilon^\sigma(p_2) \, \epsilon^\alpha(k_1) \, 
\epsilon^\beta(k_2) \, \nonumber  \\
&&\times 
[ -p_1 \cdot k_1\, C_{\mu\nu,\rho\alpha} + D_{\mu\nu,\rho\alpha}(p_1,-k_1) ]\;
[ -p_2 \cdot k_2\, C_{\mu'\nu',\sigma\beta} + D_{\mu'\nu',\sigma\beta}
(p_2,-k_2) ]\;, \\
i{\cal M}_2 &=& i{\cal M}_1 (k_1 \leftrightarrow k_2 ) \;,\\
i{\cal M}_3 &=& - \frac{\kappa^2}{8} D(s)\, B^{\mu\nu,\mu'\nu'}(p_1+p_2)\,
\epsilon^\rho(p_1) \, \epsilon^\sigma(p_2) \, \epsilon^\alpha(k_1) \, 
\epsilon^\beta(k_2) \, \nonumber \\
&&\times
[ p_1 \cdot p_2\, C_{\mu\nu,\rho\sigma} + D_{\mu\nu,\rho\sigma}(p_1,p_2) ]\;
[ k_1 \cdot k_2\, C_{\mu'\nu',\alpha\beta} + D_{\mu'\nu',\alpha\beta}
(-k_1,-k_2) ]\;,
\end{eqnarray}
where $\kappa =\sqrt{16\pi G_N}$ and $B_{\mu\nu,\rho\sigma}(k)$,
$C_{\mu\nu,\rho\sigma}$ and 
$D_{\mu\nu,\rho\sigma}(p_1,p_2)$ can be found in Ref. \cite{han}.
The propagator factor $D(s) =\sum_k i/(s-m_k^2 +i\epsilon)$, where $k$
sums over all KK levels.
After some tedious algebra the square of the amplitude, summed over final
and averaged over the initial helicities, is surprisingly
simple:
\begin{equation}
\overline{\sum} |{\cal M}|^2 = \frac{\kappa^4}{8} \; |D(s)|^2 \;
(s^4 + t^4 + u^4 ) \;,
\end{equation}
where we have taken $M_S^2 \gg s, |t|, |u|$ and in this case the propagator
factor $D(s)=D(|t|)=D(|u|)$ \cite{han}, which is given by
\begin{equation}
\kappa^2 |D(s)| = \frac{16 \pi}{M_S^4}\; \times {\cal F} \;,
\end{equation}
where the factor ${\cal F}$ is given by
\begin{equation}
\label{F}
{\cal F} = \Biggr \{ 
\begin{array}{l}
\log \left( \frac{M_S^2}{s} \right ) \;\; {\rm for}\;\; n=2 \;, \\
\frac{2}{n-2} \;\;\;\;\;\;\;\;\;\;\; {\rm for}\;\; n>2 \;.
\end{array}
\end{equation}
The angular distribution is 
\begin{equation}
\label{aa-cos}
\frac{d\sigma (\gamma\gamma \to \gamma\gamma)}{d |\cos\theta| } =
\frac{\pi s^3}{M_S^8}\, {\cal F}^2 \, \biggr [
1 + \frac{1}{8}\,(1+ 6 \cos^2\theta + \cos^4 \theta ) \biggr ] \;,
\end{equation}
where $|\cos \theta|$ is from $0$ to 1.
Since the cross section scales as $s^3/M_S^8$, which implies larger 
cross sections at higher $\sqrt{s}$.

\begin{figure}[th]
\leavevmode
\begin{center}
\includegraphics[height=4in]{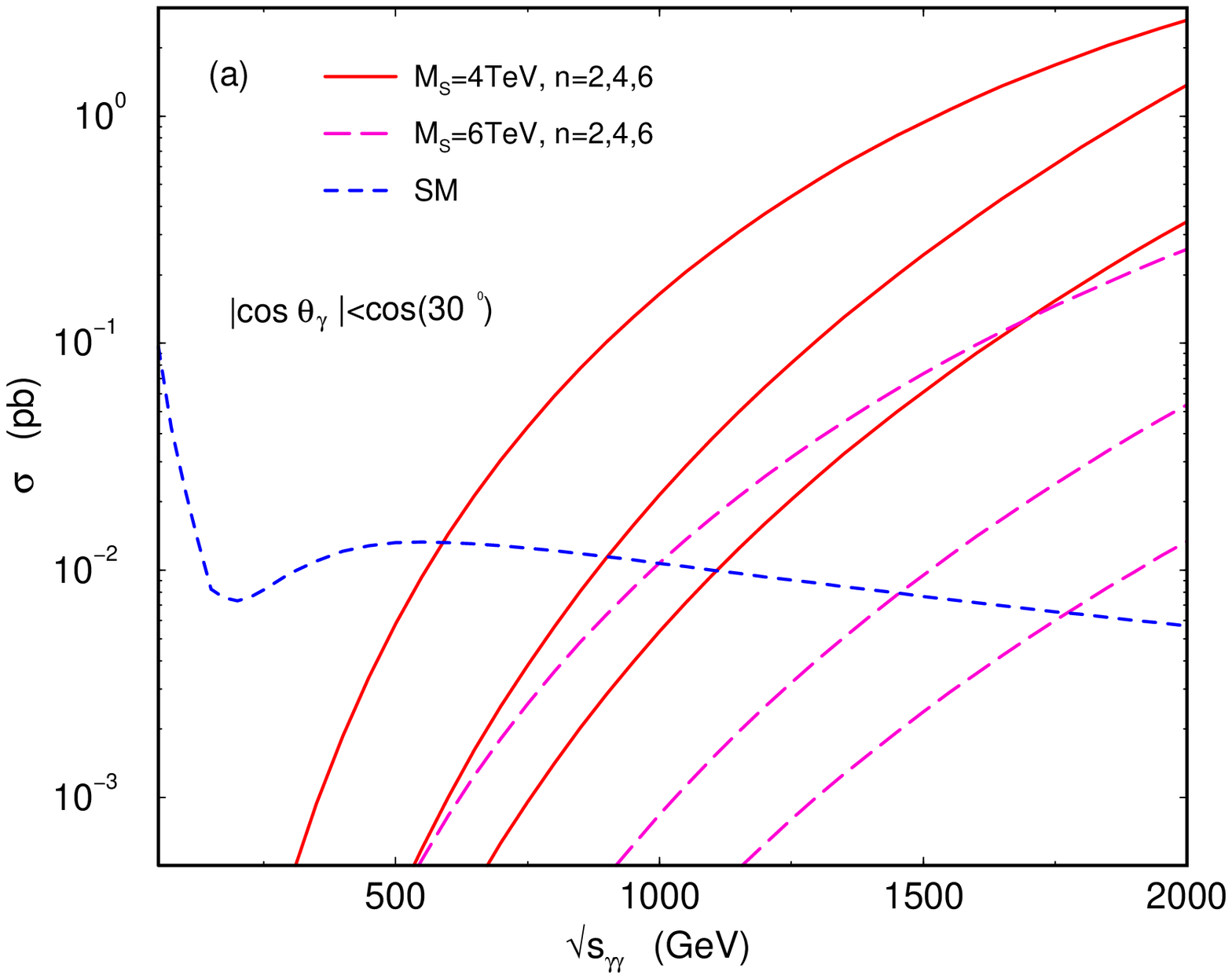}
\includegraphics[height=4in]{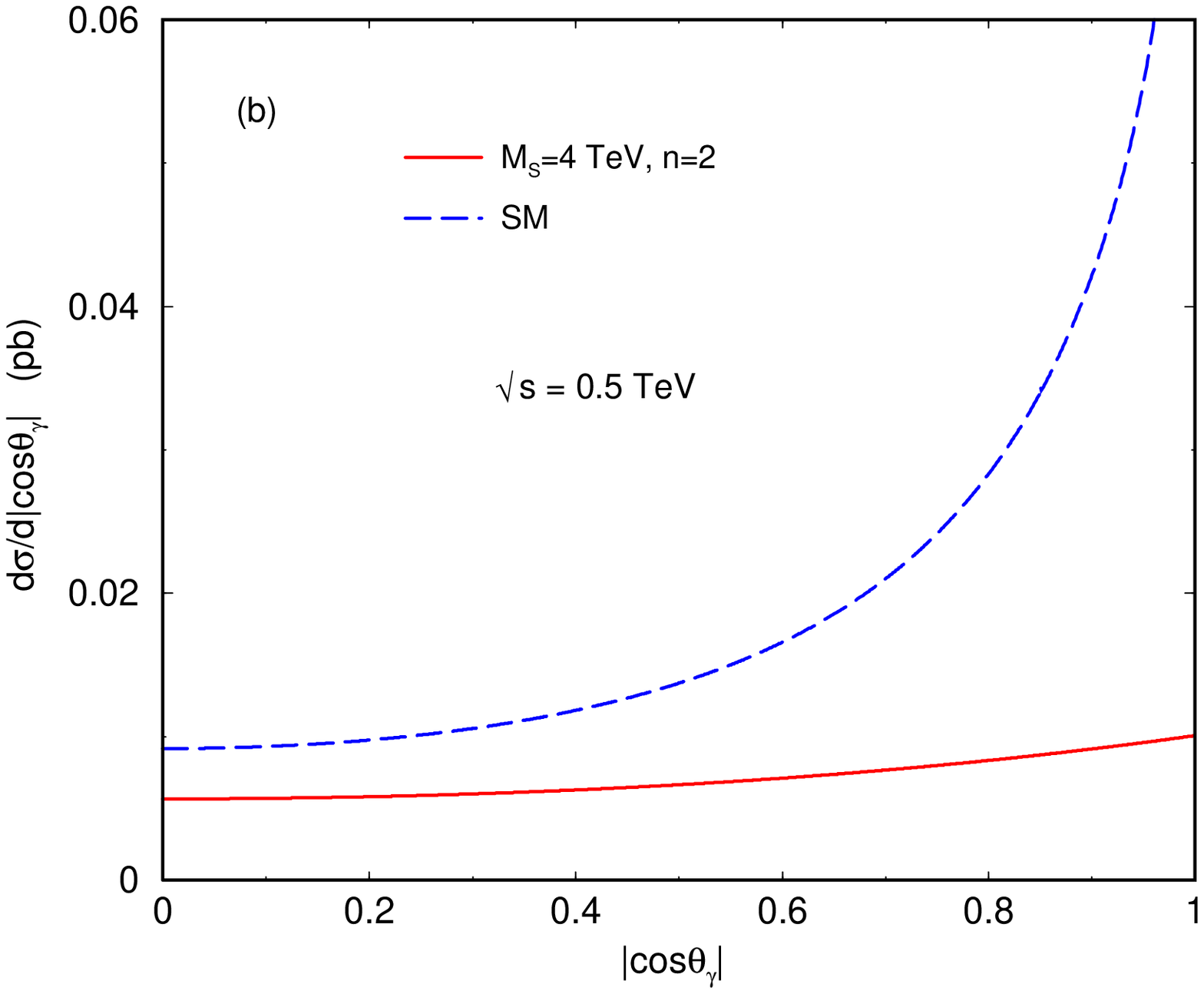}
\end{center}
\caption{
(a) The total cross sections and 
(b) the differential distribution $d\sigma/ d |\cos\theta_\gamma|$ for
$\gamma\gamma \to \gamma\gamma$ for the low scale gravity model and 
for the SM.  A cut of $|\cos \theta_\gamma| < \cos 30^\circ$ is imposed.
}
\label{fig-aa}
\end{figure}

The SM background calculation is well known and we do not repeat the
expressions here.  We used the results in Ref. \cite{photon} with the
form factors from Ref. \cite{duane}.  The process is via box diagrams
with all charged fermions and the $W$ boson in the loop.  At the low
energy, the fermion contribution dominates, but once $\sqrt{s}$ gets above
a hundred GeV the $W$ contribution becomes more important and   completely
dominates at higher $\sqrt{s}$.  We show the cross sections
in Fig. \ref{fig-aa}(a).  This SM cross section decreases gradually when 
$\sqrt{s}$ is above 500 GeV.  In contrast, the low scale gravity interactions
give a monotonically increasing cross section.
For $n=2$ and $M_S=4$ TeV the cross-over is at about $\sqrt{s}=600$ GeV.  
We notice that the signal cross section does not decrease very rapidly 
with $n$, unlike the production of real gravitons \cite{mira,ours}.

In Fig. \ref{fig-aa}(b), we show the angular distribution for the low
scale gravity and for the SM.  The signal has a relatively flat
distribution, as can be easily deduced from Eq. (\ref{aa-cos}).
The ratio of the cross section at $|\cos\theta|=0$ to that at $|\cos\theta|=1$
is only $9/16$.  On the other hand, the SM background is very steep around 
$|\cos\theta|=1$, and that is why a cut of $|\cos\theta|< \cos 30^\circ$ is
imposed to reduce the background.

\begin{figure}[th]
\leavevmode
\begin{center}
\includegraphics[height=4.5in]{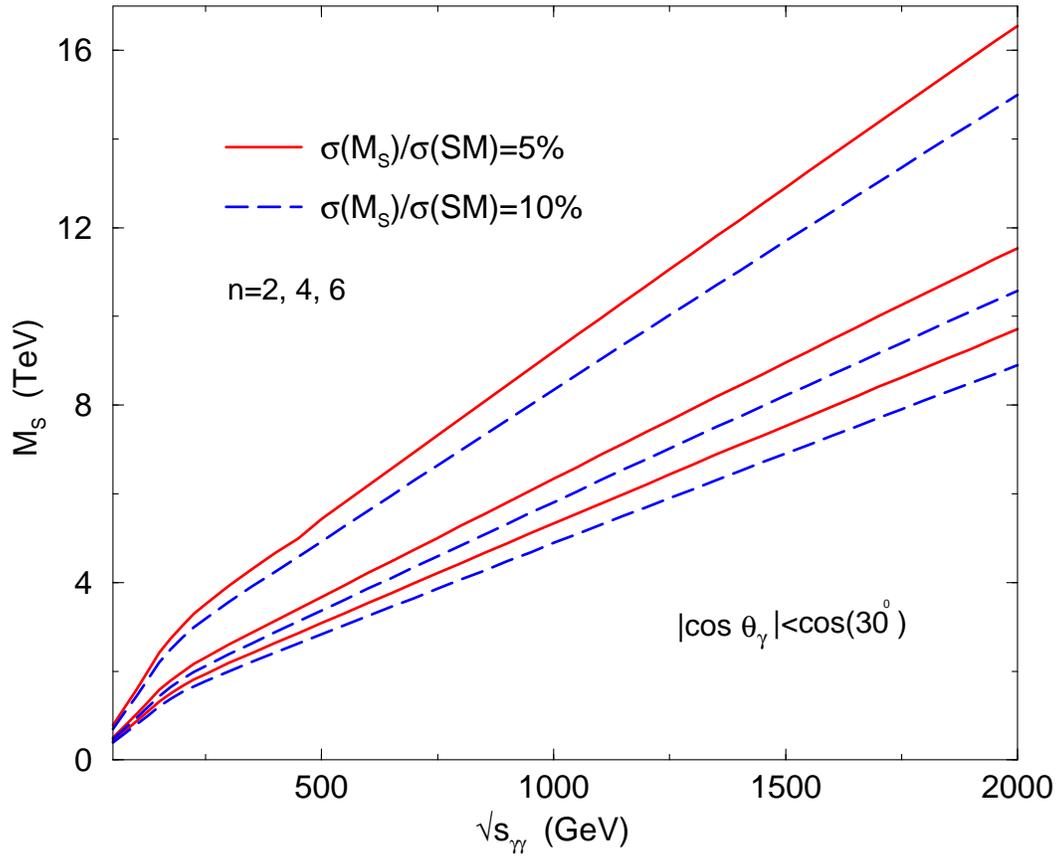}
\end{center}
\caption{ 
The sensitivity reach on $M_S$ versus $\sqrt{s_{\gamma\gamma}}$ using the
process $\gamma\gamma \to \gamma\gamma$, by requiring the signal to be 
5\% or 10\% of the SM prediction.  
A cut of $|\cos \theta_\gamma| < \cos 30^\circ$ is imposed. }
\label{fig-aa-limit}
\end{figure}

Monochromatic photon beam can be realized using the back-scatter laser 
technique \cite{telnov} 
by shining a laser beam onto an electron or positron beam.
A linear $e^+ e^-$ collider can be converted into an almost monochromatic
photon-photon collider, with a center-of-mass energy about 0.8 of the
parent $e^+ e^-$ collider and with a luminosity the same order as the
parent, i.e., as large as 50--100 fb$^{-1}$ per year.  Since the cross section
for the SM is of the order of 10 fb, so there should be enough events for
doing a counting experiment.  A 5--10 \% deviation from the SM prediction
would be at a $1.1-3.2\sigma$ level.  We use  the 5\% or 10\% deviation from
the SM as the criterion for sensitivity reach.  The sensitivity reach at
the $\gamma\gamma$ collider is shown in Fig. \ref{fig-aa-limit}.
The reach on $M_S$ is about 5--8 (4.5--7.5) times of the 
center-of-mass energy of the collider for $n=2,4,6$ using the 5\% (10\%)
deviation criterion.  
As we shall see later, the sensitivity reach at photon-photon colliders
is better than at $e^+ e^-$ and much better than at hadron colliders.

\section{Diphoton production at the Tevatron}

Diphoton production has been an interesting subject for CDF and D0.  It
can provide constraints on the $qq\gamma\gamma$ type contact interactions,
anomalous $\gamma\gamma\gamma$ and $Z\gamma\gamma$ couplings.
In the context of the low scale gravity, diphotons can be produced
via quark-antiquark and gluon-gluon annihilation into virtual gravitons and
the associated KK states.   The gluon-gluon annihilation is very similar
to the photon-photon scattering described in the last section.
The main background is the SM lowest order process: $q\bar q \to 
\gamma\gamma$.
\footnote{Since the lowest order diphoton production $q\bar q\to \gamma
\gamma$ is much larger than the box process: $gg \to \gamma\gamma$,
we shall neglect the latter in considering the SM background.}

There are two contributing subprocesses:
\begin{eqnarray}
\frac{d\sigma (q\bar q \to \gamma \gamma)}{d\cos\theta^*} &=&
\frac{1}{96\pi \hat s}\; \Biggr [
2 e^4 Q_q^4 \, \frac{1+\cos^2 \theta^*}{1-\cos^2\theta^*} +
2 \pi e^2 Q_q^2\, \frac{\hat s^2}{M_S^4} \,(1+\cos^2\theta^*) \, {\cal F} 
\nonumber \\
&& + \frac{\pi^2}{2}\, \frac{\hat s^4}{M_S^8} \, (1-\cos^4\theta^*) \,  
{\cal F}^2  \Biggr ] \;, \label{qq}
\end{eqnarray}
\begin{equation}
\frac{d\sigma (gg \to \gamma \gamma)}{d\cos\theta^*} =
\frac{\pi}{512}\, \frac{\hat s^3}{M_S^8} \, (1+ 6 \cos^2\theta^* 
+ \cos^4\theta^*) \; {\cal F}^2 \;,
\end{equation}
where the factor ${\cal F}$ is given in Eq. (\ref{F}), and the $\theta^*$ is
the scattering angle in the center-of-mass frame and $\cos\theta^*$ is from
$-1$ to 1.
In $q\bar q \to
\gamma \gamma$, the effect of graviton exchanges first occurs in the 
interference term, which only scales as $\hat s^2/M_S^4$, and 
potentially more important than the square term of $\hat s^4/M_S^8$ at 
$\hat s \ll M_S^2$.

Both CDF and D0 \cite{cdf-d0} have preliminary data on diphoton production.
We are going to use their data to constrain $M_S$.  CDF has measured
the invariant mass $M_{\gamma\gamma}$ spectrum in the region
50 GeV $<M_{\gamma\gamma}<$ 350 GeV.  However, since the data is only 
preliminary and in graphical form only, we can only use the reported
number of events in the region $M_{\gamma\gamma}>150$ GeV:
5 events are observed where $4.5\pm0.6$ are expected with an integrated
luminosity of 100 pb$^{-1}$.  This data, though without binning information,
is sufficient to place a constraint on $M_S$, because the signal for the
low-scale gravity does not appear as a peak in the $M_{\gamma\gamma}$ spectrum
but, instead, a gradual enhancement from about $M_{\gamma\gamma}\approx
150$ GeV towards higher $M_{\gamma\gamma}$.  We use the Poisson statistics
to calculate the 95\% CL upper limit to the number of {\it signal} events
$N_{95}$,
\footnote{The number of signal events is $N_{95}$ or 
less with 95\% confidence.}
using
\begin{equation}
0.95=1-\epsilon=1- \frac{e^{-(n_B + N_{95})}\; \sum_{n=0}^{n_{\rm obs}}
\, \frac{(n_B + N_{95})^n}{n !} }
{e^{-n_B}\; \sum_{n=0}^{n_{\rm obs}}
\, \frac{n_B^n}{n !} }  \;,
\end{equation}
where $n_B=4.5$ is the expected number of background events and $n_{\rm obs}
=5$ is the number of observed events.  We obtain $N_{95}=6.61$.
We then normalized our calculation to the expected number of events (=4.5) 
after imposing the same selection cuts as CDF.  With this normalization we
can then calculate $M_S$, which gives a signal of 6.61 events in excess of 
the SM prediction.  We obtain the 95\% CL lower limit on $M_S$:
\begin{eqnarray}
\mbox{Tevatron Run I:} \qquad \qquad  & &
M_S > 0.91\; {\rm TeV}\;\; {\rm for}\;\; n=2 \;\; {\rm and} \nonumber \\
&&M_S > 0.87\; {\rm TeV}\;\; {\rm for}\;\; n=4 \nonumber  \;.
\end{eqnarray}
For D0, however, the highest bin in the measured $M_{\gamma\gamma}$ spectrum 
is 80--112 GeV.  At such a low value, it is difficult to see the effect
of $(\hat s^2/M_S^4)$.  Thus, we expect the limit that would be 
obtained from the D0 data is somewhat smaller than using the CDF data.

\begin{figure}[th]
\leavevmode
\begin{center}
\includegraphics[height=4.5in]{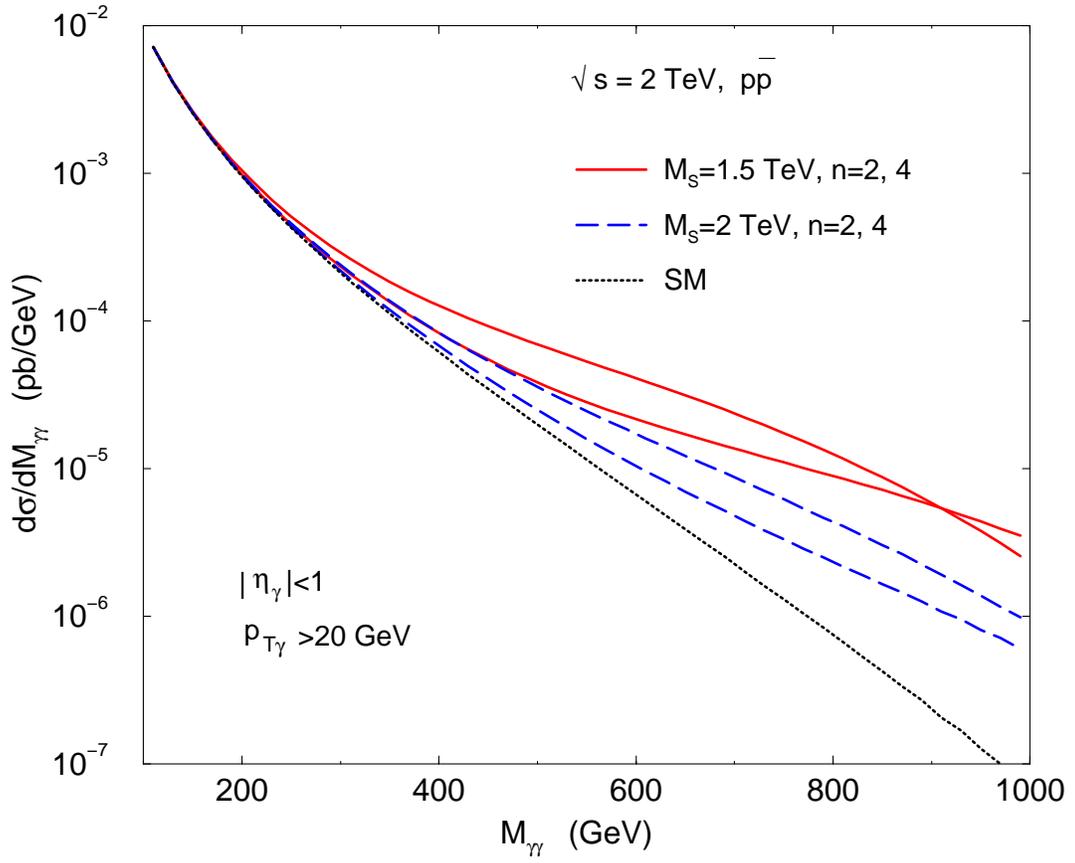}
\end{center}
\caption{ 
The differential distribution $d\sigma/dM_{\gamma\gamma}$ versus 
$M_{\gamma\gamma}$ for diphoton production at the 2 TeV Tevatron for the SM
and for the low scale gravity with $M_S=1.5,\;2$ TeV and $n=2,4$.
Cuts of $|\eta_\gamma|<1$ and $p_{T\gamma}>20$ GeV are imposed. }
\label{fig-te}
\end{figure}

Next, we estimate the sensitivity reach at the Run II of the Tevatron, 
assuming a luminosity of 2 fb$^{-1}$.  The effect of the low scale 
gravity on the $M_{\gamma\gamma}$ spectrum is shown in Fig. \ref{fig-te}.
It is easy to understand why the enhancement is more likely at the
large $M_{\gamma\gamma}$.  
To estimate the sensitivity we divide the $M_{\gamma\gamma}$ spectrum into
bins: a bin width of 100 GeV for bins in $200\;{\rm GeV} < M_{\gamma\gamma} 
< 500$ GeV, and for 500 GeV to 1000 GeV we combine it into one bin only. 
This is to make sure that each bin should have at least a few events in the SM:
see the first row of Table \ref{table-te} (we also use a selection 
efficiency of 50\%.) 
For each bin we assume the SM prediction as the number of events that 
would be observed: $n^{\rm obs}$, and we calculate the number of events
predicted by a $M_S$ and $n$: $n^{\rm th}$.  We then calculate the 
$\chi^2$ for this bin and sum over all bins, using
\begin{equation}
\chi^2(M_S,n)= \sum_{i={\rm bins}} \biggr[
2\left( n^{\rm th}_i - n^{\rm obs}_i \right ) + 2 n^{\rm obs}_i \, \ln
\left( \frac{n^{\rm obs}_i}{n^{\rm th}_i} \right )  \biggr ]\;.
\end{equation}
The $\chi^2$ then gives a goodness of the fit for the value of $M_S$ and $n$.
The larger the $\chi^2$ the smaller the probability that the corresponding
value of $M_S$ and $n$ is a true representation for the data.  
To place a 95\%CL lower limit on $M_S$ a  $\chi^2=9.49$ is needed for
4 degrees of freedom.  
The number of events in each bin for $n=2$ and $M_S=1.5-2$ TeV, and for 
$n=4$ and $M_S=1.4-2$ TeV with the corresponding $\chi^2$ are shown in 
Table \ref{table-te}.
We obtain a limit of 
\begin{eqnarray}
\mbox{Tevatron Run II:} \qquad \qquad & &
M_S>1.72\;{\rm TeV}\;\; {\rm for} \;\; n=2 \;\; {\rm and}  \nonumber \\
&&M_S>1.43\;{\rm TeV}\;\; {\rm for} \;\; n=4   \nonumber \;.
\end{eqnarray}  
We verified that the binning is not important for the limit.  We repeat the 
procedures using only one large bin from 200 to 1000 GeV, and the 95\% CL
lower limit on $M_S$ becomes 1.73 (1.38) TeV for $n=2\;(4)$.

\begin{table}[th]
\caption{ \label{table-te}
The number of events that would be observed in each bin of $M_{\gamma\gamma}$
for the SM and for the low scale gravity at the Tevatron with $\sqrt{s}=2$ 
TeV and a luminosity of 2 fb$^{-1}$.  
The $\chi^2$ is calculated 
assuming the SM prediction is what would be observed.  The
cuts imposed are: $|\eta_\gamma|<1$, $p_{T\gamma}>20$ GeV, and a selection
efficiency of 0.5 is assumed.  
}
\medskip
\begin{tabular}{c|cccc|c}
& \multicolumn{4}{c|}{bin} & \\
model & 200--300 GeV & 300--400 GeV & 400--500 GeV & 500--1000 GeV&$\chi^2$ \\
\hline
\hline
SM    & 47.68       & 11.98  & 3.65  & 1.81 & - \\
\hline
\hline
$n=2$ & & & & & \\
$M_S=2.0$ TeV & 50.27   & 14.40  & 5.53  & 4.84 & 3.80 \\
$M_S=1.9$ TeV & 50.82   & 14.92  & 5.93  & 5.54 & 5.24 \\
$M_S=1.8$ TeV & 51.53   & 15.59  & 6.46  & 6.45 & 7.33 \\
$M_S=1.75$ TeV & 51.96  & 15.99  & 6.78  & 7.01 & 8.70 \\
$M_S=1.7$ TeV & 52.47   & 16.45  & 7.15  & 7.66 & 10.37 \\
$M_S=1.6$ TeV & 53.75   & 17.61  & 8.09  & 9.30 & 14.87 \\
$M_S=1.5$ TeV & 55.49   & 19.25  & 9.38  & 11.58 & 21.72 \\
\hline
\hline
$n=4$ & & & & & \\
$M_S=2.0$ TeV & 48.24   & 12.62  & 4.22  & 2.96 & 0.64 \\
$M_S=1.9$ TeV & 48.38   & 12.77  & 4.37  & 3.28 & 0.97 \\
$M_S=1.8$ TeV & 48.54   & 12.97  & 4.55  & 3.72 & 1.49 \\
$M_S=1.7$ TeV & 48.76   & 13.24  & 4.80  & 4.38 & 2.39 \\
$M_S=1.6$ TeV & 49.07   & 13.62  & 5.15  & 5.35 & 3.89 \\
$M_S=1.5$ TeV & 49.52   & 14.16  & 5.65  & 6.87 & 6.53 \\
$M_S=1.4$ TeV & 50.20   & 14.94  & 6.40  & 9.35 & 11.30 \\
\end{tabular}
\end{table}

\section{Diphoton production at $e^+ e^-$ colliders}

We can use Eq. (\ref{qq}) with $Q_q=-1$ and multiply it by 3 to derive
the expression for $e^+ e^- \to \gamma \gamma$:
\begin{equation}
\label{ee}
\frac{d\sigma (e^+ e^- \to \gamma \gamma)}{dz} =
\frac{2 \pi}{s}\; \left(
\alpha^2  \frac{1+ z^2}{1- z^2}
+ \frac{\alpha}{4}\, \frac{s^2}{M_S^4} \, {\cal F}\, (1+z^2) 
+ \frac{1}{64}\, \frac{s^4}{M_S^8} \, {\cal F}^2 \, (1-z^4)
\right )
\end{equation}
where $z=|\cos\theta|$ is the polar angle of the outgoing photon and $z$
ranges from 0 to 1.

The four LEP collaborations have been measuring the diphoton production
$e^+ e^- \to \gamma\gamma$ \cite{lep-aa}
and using the data to constrain the deviation
from QED and generic types of contact interactions of order $1/\Lambda_n$,
$n=6,7,8$.  Since these contact interaction parameters $1/\Lambda_n$ can 
be converted from the QED cutoff parameter $\Lambda_{\pm}$, we shall
stick with the QED cutoff parameter in the following discussion.
The possible deviation from QED is usually characterized by a cutoff parameter
$\Lambda_\pm$ corresponding to a modified angular distribution:
\begin{equation}
\label{zz}
\frac{d\sigma}{dz} = \frac{2\pi \alpha^2}{s}\; \frac{1+z^2}{1-z^2} \; \left(
1 \pm \frac{s^2}{2 \Lambda_\pm^4}\, (1 - z^2 ) \right )\;,
\end{equation}
where $z = |\cos\theta|$ and ranges from 0 to 1.

Each collaboration measured the $\cos\theta$ distribution and obtained the
95\% CL limit on $\Lambda_\pm$ by varying $\eta=1/\Lambda_\pm^4$ and 
maximizing the
likelihood function.  Since each experiment has their own procedures, we
adopt a simple approach that takes their limits on $\Lambda_\pm$ and 
converts them into limits on $M_S$.  Note that in Eq. (\ref{ee}) the
third term is suppressed relative to the second term,  we can, therefore, 
just take the first and the second term, then it will look like 
Eq. (\ref{zz}).  The QED cutoff parameter $\Lambda_+$ is related to $M_S$ by
\begin{equation}
\label{convert}
\frac{M_S^4}{\cal F} = \frac{\Lambda_+^4}{2 \alpha} \;.
\end{equation}
The limits from each LEP experiment and the corresponding limits on $M_S$
are tabulated in Table \ref{table-ee}.  Note that we used only $\Lambda_+$
to calculate $M_S$.
The limits on $M_S$ is at most about 1.4 TeV for $n=2$ and about 1 TeV for
$n=4$.  The result for $n=2$ is enhanced because of the logarithmic factor
in ${\cal F}$.  Using the value of $M_S \sim 1$ TeV we
can verify the ratio of the third term to the second term in Eq. (\ref{ee})
and the third term is only about 2\% of the second term.  It justifies
the approximation that we take only the first two terms of Eq. (\ref{ee}).
So far, the treatment is rather simple. 
A better limit can be obtained by combining the data on $\eta = 
1/\Lambda_\pm^4$ from each LEP experiment. However, since some of the data 
on $\eta$ are not given in detail, we can only combine those with a central
value and an error.  We have the following available: 
(i) OPAL (183 GeV): $\eta=(1.04 \pm 1.34)\times 10^{-10}\;{\rm GeV}^{-4}$,
(ii) L3 (183 GeV): $\eta=(-0.59 
\stackrel{\scriptstyle +1.19}{\scriptstyle -1.13})
\times 10^{-10}\;{\rm GeV}^{-4}$,
(iii) L3 (161,172 GeV): $\eta=(-0.77 
\stackrel{\scriptstyle +2.83}{\scriptstyle -2.58})
\times 10^{-10}\;{\rm GeV}^{-4}$,
and (iv) DELPHI (183 GeV): $\eta=(-1.4 \pm 1.5)\times 
10^{-10}\;{\rm GeV}^{-4}$.
We combine these data and assuming they are all  gaussian we obtain
$\eta=(-0.31 \stackrel{\scriptstyle +0.74}{\scriptstyle -0.73})\times 10^{-10}
\;{\rm GeV}^{-4}$, the error of which is given in $1\sigma$.  From this $\eta$ 
the corresponding 95\% CL limit on $\Lambda_\pm$ are
$\Lambda_+ > 298$ GeV and $\Lambda_- > 279$ GeV.  
We can see that 
the combined limit on $\Lambda_+$ is still not as good as the single 
limit from ALPEH (189 GeV) or OPAL (189 GeV). 
Once the data from each LEP experiment are given we can
certainly improve the limit by combining them.  
Thus, for the present moment the best limit is from OPAL (189 GeV):
$\Lambda_+ > 345$ GeV, which converts to $M_S> 1.38\; (0.98)$ TeV for
$n=2\;(4)$.

\begin{figure}[th]
\leavevmode
\begin{center}
\includegraphics[height=4in]{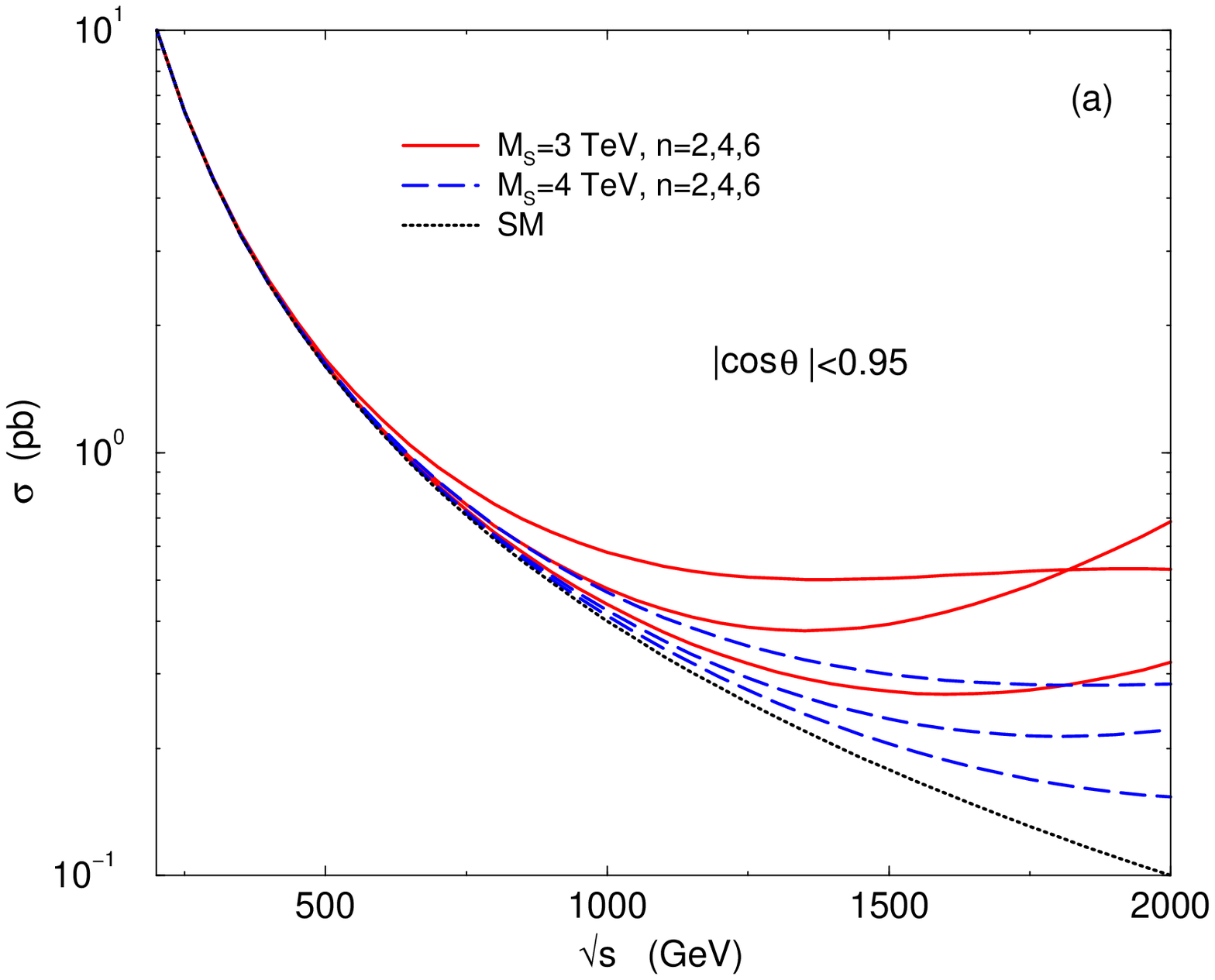}
\includegraphics[height=4in]{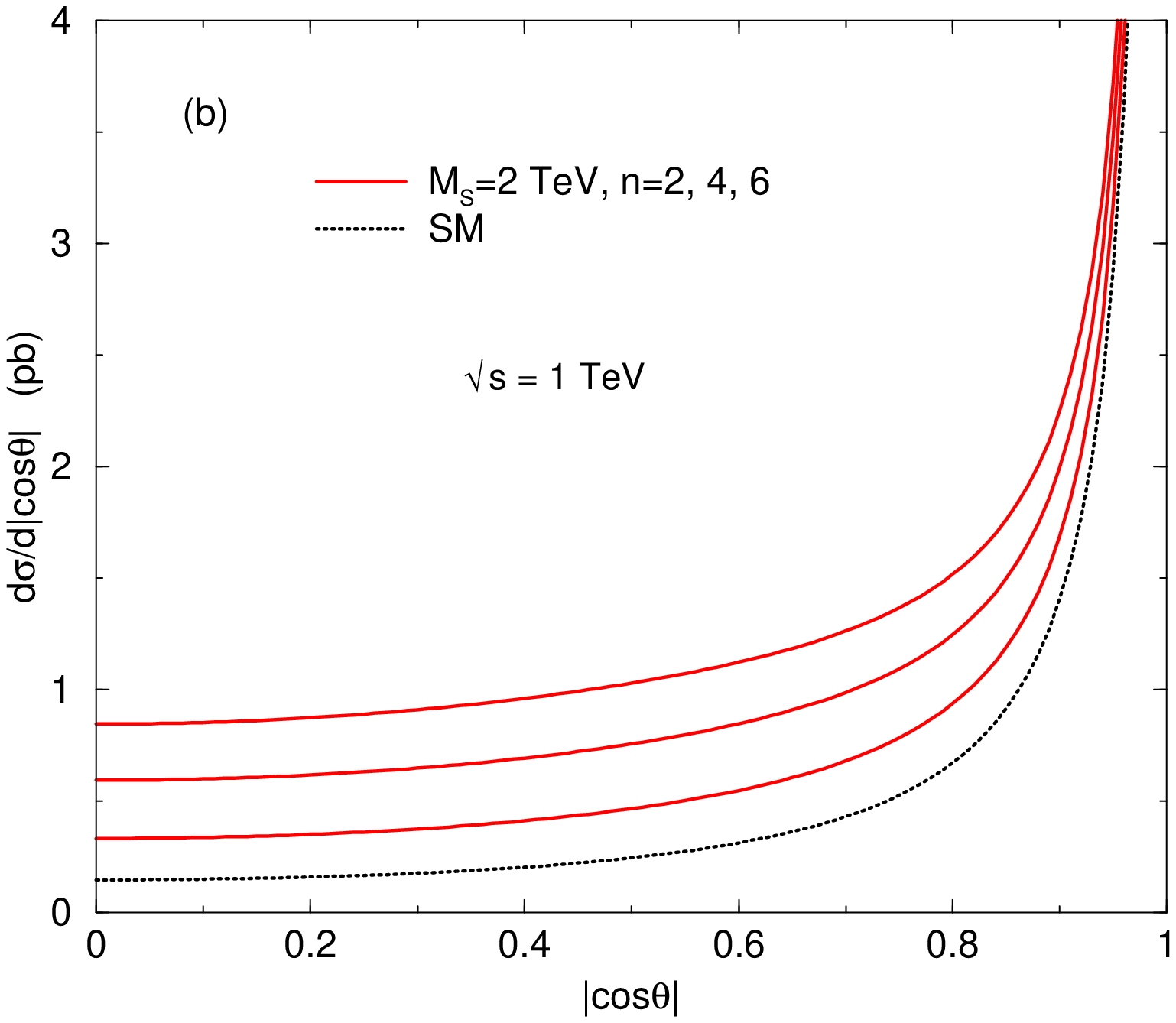}
\end{center}
\caption{ 
(a) The total cross section versus $\sqrt{s}$ and 
(b) the differential distribution $d\sigma/ d |\cos\theta |$ 
for $e^+ e^- \to \gamma\gamma$ for
the SM and for the SM plus the new gravity interactions with 
$n=2,4,6$ and $M_S$ as shown. The $|\cos\theta|<0.95$ is imposed.}
\label{fig-ee}
\end{figure}

The behavior of the new gravity interactions at higher $\sqrt{s}$ can be 
easily deduced from Eq. (\ref{ee}).  The new interaction gives rise to
terms proportional to $s^2/M_S^4$ and $s^4/M_S^8$, which get substantial
enhancement at large $\sqrt{s}$: see Fig. \ref{fig-ee}(a).  The angular
distribution also becomes flatter because in the SM the distribution scales
as 
$(1+z^2)/(1-z^2)$ whereas the terms arising from the new gravity interactions
scale as $(1+z^2)$ and $(1-z^4)$, respectively, as shown in
Fig. \ref{fig-ee}(b).

\begin{figure}[th]
\leavevmode
\begin{center}
\includegraphics[height=4.5in]{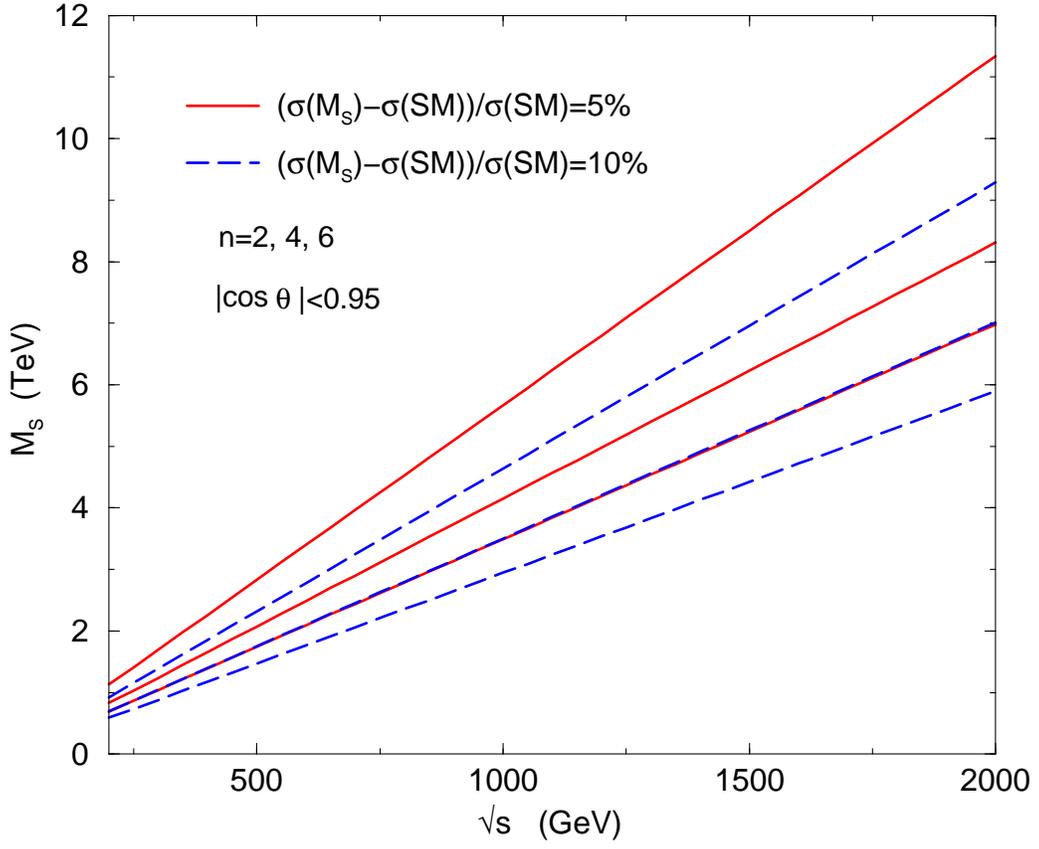}
\end{center}
\caption{ 
The sensitivity reach on $M_S$ versus $\sqrt{s}$ using the process
$e^+ e^- \to \gamma\gamma$, by requiring a 5\% or 10\% 
change from the SM prediction.  
A cut $|\cos\theta|<0.95$ is imposed.}
\label{fig-ee-limit}
\end{figure}

Here we also attempt to estimate the sensitivity reach on the cut-off scale 
$M_S$ at the future linear $e^+ e^-$ colliders.
Since the cross section is of the order of 0.1 to 1 pb for $\sqrt{s}=0.5-2$ 
TeV, it corresponds to about $10^3 - 10^4$ events for a mere yearly luminosity
of 10 fb$^{-1}$.  Thus, a 5\% (10\%) deviation from the SM prediction 
corresponds to a level of $1.6\sigma - 5\sigma$ ($3.2\sigma - 10\sigma$).
In Fig. \ref{fig-ee-limit}, we show the sensitivity reach on $M_S$ 
by requiring a 5\% or 10\% deviation from the SM prediction.
The reach on $M_S$ is about 3.5--5.5 (3--4.5) times of the $\sqrt{s}$ of 
the collider for the 5\% (10\%) criterion.

\begin{table}[t]
\caption[]{\label{table-ee}
The 95\% CL limits on the QED cutoff parameter $\Lambda_\pm$ from LEP 
experiments \cite{lep-aa} and the  corresponding 95\% limits on $M_S$
obtained using Eq. (\ref{convert}).  We show only the result of the highest
energy of each experiment whichever available.
}
\medskip
\begin{tabular}{lc|cc}
 &95\%CL limit on $\Lambda_+$ and $\Lambda_-$  & 
\multicolumn{2}{c}{95\% CL limit on $M_S$ (TeV)} \\
& &    $n=2$   &    $n=4$ \\
\hline
OPAL ($\sqrt{s}=189$ GeV): &  $\Lambda_+ >345$ GeV  &  1.38 & 0.98   \\
                           &  $\Lambda_- >278$ GeV  &       &        \\
\hline
DELPHI ($\sqrt{s}=183$ GeV): &  $\Lambda_+ >253$ GeV  & 0.97 & 0.72   \\
                             &  $\Lambda_- >225$ GeV  &      &     \\
\hline
L3 ($\sqrt{s}=183$ GeV):    &  $\Lambda_+ >262$ GeV  &  1.01 & 0.74  \\
                            &  $\Lambda_- >245$ GeV  &       &       \\
\hline
ALEPH ($\sqrt{s}=189$ GeV): &  $\Lambda_+ >332$ GeV  &  1.32 & 0.94 \\
                            &  $\Lambda_- >265$ GeV  &       &      
\end{tabular}
\end{table}

\section{Conclusions}

Diphoton production at $\gamma\gamma$, $p \bar p$, and $e^+ e^-$ colliders
provides useful channels to search for the presence of the low scale gravity
interactions, which are the effects of allowing gravity to propagate in the
extra dimensions. Photon-photon colliders are able to give the best
sensitivity reach on the cut-off scale $M_S$ of the low scale gravity model
among the three.  This is because $\gamma\gamma \to
\gamma\gamma$ can only occur via box diagrams in the SM while in $e^+ e^-$
and $p \bar p$ collisions the tree-level contributions from the SM
dominates.  In addition to the total cross section, the angular distribution
also serves as a tool to distinguish between the SM and the new gravity
interactions, as seen in Fig. ~\ref{fig-aa}(b) and Fig. \ref{fig-ee}(b).

The present limit from the LEPII diphoton data is about $M_S >1.4\;(1)$ TeV
for $n=2\;(4)$, and it is only  $M_S > 0.9$ TeV from the CDF diphoton 
$M_{\gamma\gamma}$ data.
%
The sensitivity reach in $\gamma\gamma$ collisions is about 5--8 times of
$\sqrt{s_{\gamma\gamma}}$ while it is only 3.5--5.5 times of the $\sqrt{s}$ at
$e^+ e^-$ collisions.  At the Run II of the Tevatron, the reach is only
about 1.7 (1.4) TeV for $n=2\;(4)$.

Finally, we emphasize the diphoton production at photon-photon colliders
could provide a  unique probe to the collider
signature for the model of low scale gravity.

\section*{\bf Acknowledgments}
This research was supported in part by the U.S.~Department of Energy under
Grants No. DE-FG03-91ER40674 and by the Davis Institute for High Energy 
Physics. 


\end{document}